\documentstyle[twocolumn,aps]{revtex}
\begin{document}
%
%
\title{Using electronic structure changes to map \\
the $H-T$ phase diagram of $\alpha$$^\prime$-NaV$_2$O$_5$}

\author{A.B. Sushkov and J. L. Musfeldt}
\address{Department of Chemistry,  
University of Tennessee, Knoxville, TN 37996}

\author{S.A. Crooker}
\address{National High Magnetic Field Laboratory,
MS E536, Los Alamos, New Mexico 87545}

\author{J. Jegoudez and A. Revcolevschi}
\address{Laboratoire de Physicochimie de l'Etat Solide, 
Universit\'e de Paris-Sud, B\^atiment 414, F-91405 Orsay, France}
\maketitle
%
\begin{abstract}

We report polarized 
optical reflectance studies 
of $\alpha$$^\prime$-NaV$_2$O$_5$ 
as a function of temperature (4-45 K) and magnetic field (0-60 T).
Rung directed electronic structure changes, as measured by near-infrared
reflectance ratios $\Delta$R($H$)=R($H$)/R($H$=0 T), are especially
sensitive to the phase boundaries.
We employ these changes to map out an $H-T$ phase diagram. 
Topological highlights include the observation of 
two phase boundaries slightly below $T_{SG}$, 
enhanced curvature of the 34 K phase boundary above 35 T, and,
surprisingly, strong hysteresis effects 
of both transitions with applied field. 

\end{abstract}
\pacs{PACS numbers: }

After the discovery of a low-temperature spin-gapped phase in
$\alpha$$^\prime$-NaV$_2$O$_5$, there was intense interest in
this compound as a possible spin Peierls (SP) material \cite{isobe,weiden,ueda}.
Reexamination
of the 300 K structure showed, however, that the situation
is more complex \cite{meetsma}. At room temperature, all vanadium ions
have a charge of +4.5 and form chains along the $b$ axis. Chains are 
connected by  V-O-V rungs. 
The result is a  1/4-filled 2-leg ladder material 
with low-dimensional 
chain character and 1 spin per rung \cite{smolinski,horsch}. 
The spin gap (SG) opens at $T_{SG}$ = 34 K, as evidenced by an
isotropic drop in the susceptibility \cite{isobe,weiden,ueda}.
The opening of the spin gap 
is accompanied by unit cell doubling in the $a$ and $b$ directions, 
and quadrupling in 
the $c$ direction
\cite{vansmaalen,nakao,isobe2,fujii,seo,mostovoy,thalmeier,nishimoto1,vasilevprl,koppen,schnelle}.
According to L\"udecke {\em et al.},
the low temperature phase of $\alpha$$^\prime$-NaV$_2$O$_5$
also displays alternating modulated and unmodulated ladder chains 
\cite{vansmaalen}, although Nakao {\it et. al.} report a fully zigzig
ordered structure, with variable temperature charge disproportionation
\cite{nakao}. 
Early NMR experiments pointed out the importance of charge ordering (CO) in
this material, as evidenced by both V$^{4+}$ and V$^{5+}$ resonances
\cite{ohama2}.
Zigzag CO is the leading candidate for the
low-temperature charge arrangement \cite{nakao,mostovoy}.
More recent NMR experiments have
discovered the advent of the charge-ordered phase in
$\alpha$$^\prime$-NaV$_2$O$_5$ at $T_{CO}$ = 37 K,
slightly above the 34.7 K lattice distortion and spin gap formation
\cite{fagot}. Early thermal expansion measurements also
discerned two transitions \cite{thermal}. 
The detailed interaction
between charge and magnetic ordering in
$\alpha$$^\prime$-NaV$_2$O$_5$ is not understood at this time,
although, because of
the larger energy scale, CO is expected to dominate. 
Further, the field dependence of $T_{SG}$ (on  a field vs. temperature ($H-T$)
phase diagram) 
is anomalous \cite{cross,bulaevskii} and does not follow
the expected $H^2$ behavior for a SP transition, at least at low
applied fields \cite{schnelle,fertey,uf,paul}. The
nearly vertical field dependence of $T_{SG}$ and the interplay of
CO and magnetic effects 
have been cited as reasons why 
$\alpha$$^\prime$-NaV$_2$O$_5$ should not be considered a true SP material.
 
The electronic structure of $\alpha$$^\prime$-NaV$_2$O$_5$ has
attracted a great deal of attention 
\cite{golubchik,damascelli,popova2,long2,damascelli2,presura1,presura2}, as
have other physical properties
\cite{vasilevprl,koppen,fertey,sherman,smirnov99,luther,nojiri,vasilevesr,konstantinovic,postolache,smirnov1}.
The optical conductivity
is strongly polarized, as expected for such an anisotropic
material; excitations in the rung direction ($a$) are much stronger
than those in the chain direction ($b$). Electronic features
are observed near 1 eV, between 3 and 4 eV, and above 4 eV in both
polarizations. The 1 eV structure in the $a$ direction has
received the most attention and
is currently assigned as charge transfer excitation between  
V$^{4+}$ and V$^{5+}$ on opposite sides of a rung \cite{presura2}.  
That changes in electronic structure are sensitive probes
of phase boundaries 
was established by previous work on CuGeO$_3$ and
$\alpha$$^\prime$-NaV$_2$O$_5$, where integrated intensities of
reflectance and transmittance ratio features, as a function of temperature or
magnetic field, 
display ``breakpoints" at a phase boundary \cite{long2,long1,zeman}.
These inflection points provide
a useful
tool to locate a transition in $H-T$ phase
space. In CuGeO$_3$, the phase diagram obtained in this manner is
consistent with that obtained with other techniques \cite{long2,long1,zeman}.
At the same time, the detailed spectral
features provide microscopic information on the changes which occur
at a phase boundary.
In the past, the
challenge of accessing the high field phase 
has prevented general extension of this
technique to $\alpha$$^\prime$-NaV$_2$O$_5$.

In order to provide further information on the
electrodynamics of quarter-filled magnetic oxides,
we have carried out polarized optical reflectance
measurements to probe the high field
response of $\alpha$$^\prime$-NaV$_2$O$_5$. 
We use these data to
generate an $H-T$ phase diagram which shows
unusual topology and a number of important differences from traditional
SP behavior.

High quality single crystals of  
$\alpha$$^\prime$-NaV$_2$O$_5$ were grown in Orsay by 
the flux method \cite{crystal}. Typical crystal dimensions were
1.65 $\times$ 2.8 $\times$ 0.25 mm.
The high-field optical reflectance measurements were performed in the 60 T
Long-Pulse magnet
at the National High Magnetic Field Lab in Los Alamos, NM. This magnet is
powered by a 1.4 GVA motor-generator which provides a very long
(2 second), user-definable magnet pulse. We elected to use the
flat-top field profile, in which the magnet is held at
maximum field (60 T) for a full 100 ms. The sample resides in a
vacuum jacket in the bore of the magnet, and data were taken in the range
4 to 45 K, concentrating in the regime around the 34 K phase transition.
The temperature was held constant for each shot; the
sample was thermally cycled to above 40 K after each magnet pulse to avoid
possible hysteresis/fatiguing  effects.
White light was coupled to the sample via a
single 600 $\mu$m optical fiber, and thin-film, in-situ linear polarizers
selected the desired polarization along the $a$- (rung) or $b$- (chain)
direction. The reflected light was collected by another optical fiber,
dispersed in an optically fast 0.3 m spectrometer, and collected by a
high-speed CCD camera.  Because the optical spectra of
$\alpha$$^\prime$-NaV$_2$O$_5$ is modified
in broad bands, a 150 line/mm grating was employed.
We covered the wavelength range from 350 to 1200 nm.
Full spectra
were collected every 2.1 ms throughout the entire 2 second magnet pulse,
providing for  complete field-dependence in a single 
shot at fixed field \cite{note10}. A calibrated coil, mounted on the probe, measured
the applied field, which was applied perpendicular to
the $ab$ plane.

Small field-induced changes in the polarized optical reflectance 
were studied by calculating
reflectance ratios,   
$\Delta$R($H$) = R($H$)/R($H$=0 T), at a number of
fixed temperatures.
Reflectance ratios were measured in 1 or 1.5 K steps around $T_{SG}$ (34 K) 
and at larger temperature intervals further away from the transition.
We quantified 
these reflectance ratio changes with applied field
by integrating the spectral
area. 
Inflection points in the integrated area 
were used to identify
phase boundaries in $H-T$ space.
Error bars on the location of the high-field phase
boundaries vary depending on
the sweep; we estimate the largest to be on the order of
$\pm$ 3 T on the 30 K point
near 60 T; other error bars are $\pm$ 2 T.

Figure 1 displays the rung-polarized ($a$ axis) 
near-infrared reflectance ratios 
of $\alpha$$^\prime$-NaV$_2$O$_5$ with applied field.
As in previous temperature sweep data, the
electronic structure is modified in wide bands, with
the largest changes occurring in the vicinity of the 1 eV
excitation \cite{long2}. That the 1 eV excitation is
sensitive to an applied magnetic field
is related to the charge transfer from vanadium centers across the V-O-V
rung, which also involves a shift of the spin to a new site. 
Chain-polarized
and higher energy rung-polarized
spectral signatures were also observed at the
phase boundaries; 
we concentrate our analysis and
discussion on the rung-directed near-infrared data because the effects are
more pronounced.

In order to quantify the spectral changes in
$\alpha$$^\prime$-NaV$_2$O$_5$ with magnetic field, we calculated the
spectral area in the wavelength range of interest and plotted it
vs. applied field. Representative data
are shown in Fig.~2.
The most interesting spectral changes are 
in the vicinity of the 34 K transition, where  
there are large hysteresis loops in the integrated
area vs. field data below $T_{SG}$. 
That inflection points in the optical response are different
on the upsweep compared to the downsweep
indicates first-order behavior.
Data collected above the 34 K transition temperature does not
display hysteresis (Fig.~2). 
In contrast, CuGeO$_3$
shows second-order character over the majority of $H-T$ phase space
with first order behavior only around the critical field
($H_c$=12.5 T) at low temperature, where the spin
gap is fully open and the lattice distortion is most pronounced \cite{hase1,hase2}.
Previous investigations of $\alpha$$^\prime$-NaV$_2$O$_5$ generally report
second-order behavior \cite{ravy}, although these efforts have not explored
the $H-T$ diagram in the field regime reported here. The thermal expansion
data are an exception in that two transitions are observed \cite{thermal}.

Figure 3 displays the $H-T$ phase diagram for
$\alpha$$^\prime$-NaV$_2$O$_5$ derived from an analysis of the
field dependent optical properties. 
The field dependence of $T_{SG}$ is nearly
vertical close to 34 K, softening
only above 35 T. 
Note that previous 28 and 30 T determinations of the transition
temperature are in general agreement with this diagram \cite{uf,paul,long2}.
Two phase boundaries are clearly
observed below $T_{SG}$ in $\alpha$$^\prime$-NaV$_2$O$_5$.
That there are indeed two phase boundaries  
is supported by the spectral area vs. field data in 
Fig.~2.
Here, two distinct inflection points
are seen in both up- and down-field sweeps; the position of these
breakpoints changes with temperature, indicating that they are characteristic
of $\alpha$$^\prime$-NaV$_2$O$_5$ and not an experimental artifact
\cite{note11}.
The low-field boundary in Fig.~3
separates two phases directly below $T_{SG}$;  it 
saturates near 33 T.
This structure is reminiscent of, but different than, the critical field ($H_c$) 
boundary that separates dimerized and incommensurate phases 
in traditional SP materials \cite{bray}.
The optical signature of the 33 T boundary
weakens rapidly away from $T_{SG}$, as indicated on the phase diagram,
making the distinction between the two
phases short-lived and limiting the range of phase space available
for study \cite{explain}.
In contrast, the optical signature of the
high field phase boundary 
remains strong within the range of our
investigation (up to 60 T).
The curvature is weaker than that in CuGeO$_3$ 
\cite{hase1,hase2}, but stronger than that previously reported
for $\alpha$$^\prime$-NaV$_2$O$_5$ \cite{schnelle,uf}. 
Based upon the three highest field data points
and the work of Bulaevski and
Cross \cite{cross,bulaevskii} which predicts $H^2$
dependence to the phase boundary, we extract $\alpha$$\sim$0.1.
The first-order behavior of this (and the 33 T) phase boundary is  
unexpected and 
different from CuGeO$_3$ \cite{hase1,hase2}, although recent theoretical
work proposes a framework under which hysteresis 
might be understood \cite{uhrig}.

In summary, we report polarized 
optical reflectance studies 
of $\alpha$$^\prime$-NaV$_2$O$_5$ between 350 and 1200 nm 
as a function of temperature (4-45 K) and applied magnetic field (0-60 T).
Using inflection points in the $a$-axis 
near-infrared reflectance ratio
data, we have identified high-field phase
boundaries at a number of different temperatures 
and mapped out an $H-T$ phase diagram. 
Highlights include the observation of 
two phase boundaries slightly below $T_{SG}$, 
enhanced curvature of the 34 K phase boundary above 35 T, and,
surprisingly, strong hysteresis effects 
of both transitions with applied field. 
We hope these experiments will spur complementary magnetization
and NMR studies of $\alpha$$^\prime$-NaV$_2$O$_5$ in the future.

\vspace{.3in}

\centerline{\bf {Acknowledgements}} 

We are grateful for financial support from
the Materials Science Division, Basic Energy Sciences at the U.S. Department
of Energy under Grant \# DE-FG02-99ER45741.
The National High 
Magnetic Field Laboratory in Los Alamos, NM, is  
supported by the National Science Foundation,
the State of Florida, and the Department of
Energy. We thank Bill Moulton for useful discussions.

\centerline{FIGURES}                                                           

Fig. 1. Near-infrared reflectance ratio ($\Delta$R)
spectra of $\alpha$$^\prime$-NaV$_2$O$_5$ at
31 K polarized in the rung ($a$) direction. The $\Delta$R spectra
shown here are presented at three representative fields: 60 T/0 T, 45 T/0 T, and
0 T/0 T, illustrating the different
optical response in the three 
regimes of the $H-T$ phase diagram at this temperature. \\
\\

Fig. 2. Integrated spectral area of $\alpha$$^\prime$-NaV$_2$O$_5$
vs. applied magnetic field
at four different temperatures below $T_{SG}$ and one temperature above
$T_{SG}$. Solid arrows indicate breakpoints; dashed arrows denote
the field sweep direction. \\
\\

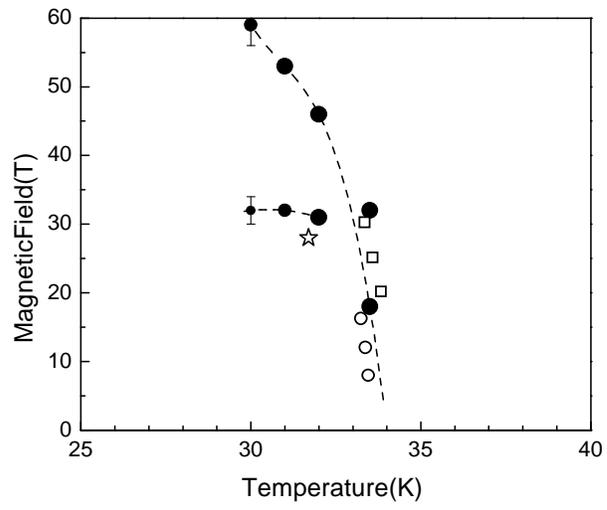
Fig. 3. Solid symbols: $H-T$ phase diagram of
$\alpha^\prime$-NaV$_2$O$_5$, as obtained
from an analysis of inflection
points in the reflectance ratio data on the upsweep of
the magnet. 
Symbol size is intended
to allow visualization of transition strength. Open star: 28 T data as
determined by Long {\it et. al.} in Ref. \cite{long2}; 
open squares:  data
as determined by Bompadre {\it et. al.} in  Ref. \cite{uf}; open
circles: data as determined by Schnelle {\it et. al.} in
Ref. \cite{schnelle}. Dashed lines guide the eye.  \\
\\


\begin{references}
\bibitem{note10} There are three parts to each shot, each with different field
vs. time profiles. Since we collect spectra ever 2.1 ms, different
field resolutions are therefore obtained over the pulse profile.
\bibitem{explain}
Possible explanations
for this progressive weakening 
include an extremely wide low-temperature hysteresis (making the transition
impossible to follow with only 60 T pulses) or
domination of lattice dimerization effects
by CO. Clearly, the decay of this boundary is of interest in future work.
\end{references}
\end{document}